\pdfoutput=1

\documentclass[12pt, a4paper]{article}

\usepackage{xcolor}
\usepackage{framed}
\usepackage{amsmath}
\usepackage{amsfonts}
\usepackage{amssymb}
\usepackage{graphicx, rotating}
\usepackage{epsfig}
\usepackage{latexsym}
\usepackage{graphicx}
\usepackage{color}
\usepackage{amsmath,bm,amssymb}
\usepackage{cite}
\usepackage{slashed}
\usepackage{hyperref}
\usepackage{epstopdf}
\AppendGraphicsExtensions{.tif}
\hypersetup{colorlinks=true, citecolor=bluscuro, linkcolor=black, urlcolor=bluscuro}
\definecolor{rossos}{cmyk}{0,1,1,0.55}
\definecolor{bluscuro}{rgb}{0.15, 0.2, .85}
\definecolor{bluchiaro}{cmyk}{1,.3,0.,0.1}

\newcommand{\be}{\begin{eqnarray}}
\newcommand{\ee}{\end{eqnarray}}

\usepackage{amsmath}
\numberwithin{equation}{section}
\def\wh{\widetilde{h}}
\def\ba{\begin{eqnarray}}
\def\ea{\end{eqnarray}}

\def\bq{\begin{quote}}
\def\eq{\end{quote}}

 at 10truept

\newcommand{\beq}{\begin{equation}}
\newcommand{\eeq}{\end{equation}}
\newcommand{\beqa}{\begin{eqnarray}}
\newcommand{\eeqa}{\end{eqnarray}}
\newcommand{\bea}{\begin{eqnarray}}
\newcommand{\eea}{\end{eqnarray}}

\newcommand{\mpl}{M_{Pl}}

\def\ltap{\ \raise.3ex\hbox{$<$\kern-.75em\lower1ex\hbox{$\sim$}}\ }
\def\gtap{\ \raise.3ex\hbox{$>$\kern-.75em\lower1ex\hbox{$\sim$}}\ }
\def\gl{\ \raise.5ex\hbox{$>$}\kern-.8em\lower.5ex\hbox{$<$}\ }
\def\roughly#1{\raise.3ex\hbox{$#1$\kern-.75em\lower1ex\hbox{$\sim$}}}

\def\d{{\rm d}}
\def\mpl{M_{\rm pl}}

\newcommand{\arXiv}[2]{\href{http://arxiv.org/pdf/#1}{{\tt [#2/#1]}}}
\newcommand{\arXivold}[1]{\href{http://arxiv.org/pdf/#1}{{\tt [#1]}}}

\begin{document}

\begin{titlepage}
\begin{flushright}
\end{flushright}
\vspace{0.1in}

\begin{center}
{\Large\bf\color{black}
Clockwork Inflation 
}\\
\bigskip\color{black}
\vspace{1cm}{
{\large Alex ~Kehagias$^{a,b}$ and Antonio ~Riotto$^c$}
\vspace{0.3cm}
} \\[7mm]
{\it {$^a$\, Physics Division, National Technical University of Athens\\ 15780 Zografou Campus, Athens, Greece}}\\
{\it {$^b$\, Theoretical Physics Department, CERN, CH-1211 Geneva 23, Switzerland}}\\
{\it $^c$ {Department of Theoretical Physics and Center for Astroparticle Physics (CAP)\\ 24 quai E. Ansermet, CH-1211 Geneva 4, Switzerland}}\\
\end{center}
\bigskip

\vspace{1cm}

\begin{abstract}
\noindent
We investigate the recently proposed clockwork mechanism delivering light degrees of freedom with suppressed interactions and show, with various examples,  that it can be efficiently implemented in inflationary scenarios to generate flat inflaton potentials and small density perturbations without fine-tunings.
We also study the clockwork graviton in de Sitter and,  interestingly, we find that the corresponding clockwork charge is site-dependent. As a consequence, 
the amount of tensor modes is generically suppressed with respect to the standard cases where the clockwork set-up is not adopted. This point can be made a virtue in resurrecting models of inflation which were supposed to be ruled out because of the excessive amount of tensor modes from inflation. 
\end{abstract}
\bigskip

\end{titlepage}

\baselineskip=18pt
\section{Introduction \label{sec:intro}} 
The clockwork mechanism \cite{kor,kr}  allows to explain the presence of light degrees of freedom  with highly  suppressed interactions in theories where there are no small parameters to start with. A general theory of  the clockwork mechanism valid for scalars, fermions, gauge bosons, and gravitons has been recently proposed in 
Ref. \cite{gm}.  Let us briefly show how it operates for scalars and consider
a theory endowed with  a global U(1)$^{N+1}$ spontaneously broken at the scale $f$. The  degrees of freedom at energies smaller than $f$ are the $N+1$ Goldstone bosons $\pi_i$
\beq
U_i (x)= e^{i\pi_i (x)/f}, ~~~i=0, \cdots, N. 
\eeq 
The $\pi_i$ fields  transform by a phase under the corresponding Abelian factor U(1)$_i$. Suppose now that the low-energy description of the theory
is described by the Lagrangian

\begin{eqnarray}
{\mathcal L} &=& -\frac{f^2}{2}\, \sum_{i=0}^{N} \partial_\mu U_i^\dagger\, \partial^\mu U_i +\frac{m^2f^2}{2}\, \sum_{i=0}^{N-1} \left( U_i^\dagger\, U_{i+1}^q + {\rm h.c.} \right)\nonumber\\
&=&-\frac{1}{2} \sum_{i=0}^N (\partial \pi_i)^2 +\frac{m^2}{2} \sum_{i=0}^{N-1}
\left( \pi_i  -q\, \pi_{i+1}\right)^2 + {\mathcal O} (\pi^4)\nonumber\\
&=&-\frac{1}{2} \sum_{i=0}^N (\partial \pi_i)^2+\frac{1}{2} \sum_{i,j=0}^N \pi_i \, M^2_{\pi \, ij}\, \pi_j,
\label{lag1}
\end{eqnarray}
where the presence of the explicit mass terms breaks soflty the symmetry U(1)$^{N+1}$ down to a single U(1). The square mass matrix is given by

\beq
M^2_\pi = m^2
\begin{pmatrix}
1 & -q & 0 & \cdots &  & 0 \cr
-q & 1+q^2 & -q & \cdots &  & 0 \cr
0 & -q & 1+q^2 & \cdots & & 0 \cr
\vdots & \vdots & \vdots & \ddots & &\vdots \cr
 & & & & 1+q^2 & -q \cr
 0 & 0 & 0 &\cdots & -q & q^2
\end{pmatrix} .
\label{pimass}
 \eeq
 In the mass eigenstate  basis $\phi_i$ ($i=0,\, ...,N$)
\beq
\pi = O\, \phi, ~~~~~~O^T M_\pi^2 O = {\rm diag}\, (m^2_{\phi_0},\dots , m^2_{\phi_N}),
\eeq
where $O$ is a real orthogonal matrix,  the eigenvalues are given by
\beq
m^2_{\phi_0} =0,~~~m^2_{\phi_k} =\lambda_k\, m^2, ~~~ \lambda_k \equiv q^2 + 1 -2q \cos \frac{k\pi}{N\! +\! 1}, ~~~~~ k=1, \cdots, N.
\label{mass}
\eeq
The elements of the rotation matrix $O$ are given by 
\beq
O_{i 0} = \frac{{\cal N}_0}{q^i}, ~~~ 
O_{i k} = {\cal N}_k \left[ q \sin \frac{i k\pi}{N\! +\! 1}-  \sin \frac{(i +1) k\pi}{N\! +\! 1} \right],~~~ i =0, .. , N;~~k =1, \cdots , N,
\label{rotation}
\eeq
and
\beq
{\cal N}_0 \equiv \sqrt{\frac{q^2-1}{q^2-q^{-2N}}} \, , ~~~~ {\cal N}_k \equiv \sqrt{\frac{2}{(N\! +\! 1)\lambda_k}}.
\eeq
The key point of the clockwork mechanism is now that the massless eigenstate $\phi_0$ is  coupled to the rest of the fields in the theory with a coupling which is
suppressed by $O_{i0}\sim q^{-i}$. In particular, if the rest of the degrees of freedom in the matter sector couples only to the $N$-th pion $\pi_N$, the state $\phi_0$ couples to them with a suppressed coupling scaling like $q^{-N}$. If $N$ is large and $q>1$, then the coupling is efficiently suppressed.

In the case in which the number of copies is very large, it has been pointed out that there exists also a five-dimensional 
continuum limit of the clockwork mechanism \cite{gm}. It is achieved by introducing a dilaton field $S$ in a five-dimensional braneworld with the fifth dimension compactified on $S_1/{\mathbb Z}_2$. The corresponding action reads

\begin{eqnarray}
{\cal S}&=&\int \d ^4x\d y\sqrt{-g}\left\{\frac{M_5^3}{2}
\left(R-\frac{1}{3}\partial_M S\partial^M S+4k^2 e^{-\frac{2}{3}S}\right)\right.\nonumber\\
&-&\left.\frac{e^{-\frac{1}{3}S}}{\sqrt{g_{55}}}
\left[\delta(y)V_0+\delta(y-\pi R)V_\pi\right]\right\},
\end{eqnarray}
where $k^2$ characterizes the negative vacuum energy in the bulk, $R$ is the radius of the fifth dimension, $M_5$ is the fundamental scale in the bulk, and $V_0$ and $V_\pi$ are tensions on the brane satisfying the relation $V_0=-V_\pi=-4 k M_5^3$. The corresponding metric
is found to be

\beq
\d s^2=e^{\frac{4}{3}k|y|}\left(\d y^2+\eta_{\mu\nu}\d x^\mu\d x^\nu\right),
\eeq
with $\eta_{\mu\nu}$ the flat Minkowski metric. In this picture hierarchies are produced on the $y=\pi R$ brane and  the discrete suppression   factor $q^{-N}$ is replaced in the   continuum with $e^{-k\pi R}$.

The goal of this note is to show that the clockwork mechanism can be  adopted in inflationary theories to efficiently generate 
flat inflaton potential sustaining a de Sitter phase as well as small masses and couplings to match the small amount of observed scalar perturbations. We will present in section 2 various examples of such ability from the four-dimensional discrete perspective. 
In section 3 we will study the phenomenon of inflation from the five-dimensional continuum perspective and show that the amount of 
clockworking producing small masses/couplings depends on the Hubble rate during inflation.
Maybe more interestingly, in section 4 we show that, within the clockwork set-up, the clockwork charges of the gravitons are site-dependent
and 
the amount of tensor modes generated during inflation is suppressed with respect to the standard scenario due to the fact that tensor modes are intrinsically bulk degrees of freedom.
Section 5 contains our conclusions.

\section{Clockwork inflation: the four-dimensional discrete perspective}
In this section we show how to exploit the  clockwork theory in inflation. The clockwork set-up is suitable to get either small masses (compared to the fundamental mass scale of the problem) or small couplings (compared to  couplings of order unity). This is exactly what is needed during inflation in order to get   the right amount of density fluctuations.
The  comoving curvature perturbation $\zeta$ in the flat gauge is \cite{lr}
\be
\label{cp}
\zeta=\left(\frac{H}{\dot\phi}\right)_*\,\delta\phi\sim\left(\frac{H}{\mpl\sqrt{\epsilon}}\right)_*,
\ee
where the subscript ${}_*$ indicates that quantities should be computed  at the epoch of Hubble radius 
exit for the comoving scale $k=a H$, $\phi$ is the inflaton field, $H$ is the Hubble rate during inflation,   $M_{\rm pl}$ is the reduced Planck mass,   and  one has to remember that observables scales in our current universe correspond to the last 60 $e$-folds or so before the end of inflation.  Dots indicate differentiation with respect to time and 

\be
\epsilon=-\frac{\dot H}{H^2}\simeq\frac{\mpl^2}{2}\left(\frac{V'}{V}\right)^2
\ee
is one of the slow-roll parameters.
The observed perturbations are matched if   

\be
\label{cobe}
\left(\frac{V^{3/2}}{\mpl^3 V'}\right)_*
\simeq 5.3\cdot 10^{-4}.
\ee

\subsection{Large field models of inflation}
To illustrate the advantages of the clockwork set-up in producing flat potential during inflation, 
let us consider  the class of large field models of inflation. The simplest model of inflation is given by a linear  potential

\beq
V(\phi)=m^3\phi.
\eeq
We know that the slow-roll conditions are attained when  $\phi\gg M_{\rm pl}$  and that the density perturbations are given by \cite{lr}

\beq
\zeta\sim\left(\frac{m}{M_{\rm pl}}\right)^{3/2}\sim 10^{-5},
\eeq
for $m\sim 10^{15}$ GeV $\ll M_{\rm pl}$. In the clockwork scenario, we can assume that there are $N+1$ copies of the inflaton fields
and the potential is

\beq
V(\pi_1,\cdots,\pi_N)=\frac{M_1^2}{2}\sum_{i=0}^{N-1}\left(\pi_i-q \,\pi_{i+1}\right)^2+M_2^3\pi_N,
\eeq
where $M_2\ll M_1$ (say smaller by a factor of 10), but  both close to the fundamental scale. The first piece of the potential is invariant under a shift symmetry

\beq
\pi_i\to \pi_i +\frac{1}{q^i},
\eeq
which is broken by the last term in the potential. 
Upon diagonalization of the mass matrix (which is not altered by the presence of the linear term)  and going to energy (to be identified with the Hubble rate $H$) much smaller than $M_2$, the lightest mass eigenstate state $\phi_0$ will have a potential

\beq
V(\phi_0)=\frac{M_2^3}{q^N}\phi_0. 
\eeq
Taking for instance $M_2\sim 10^{-1}M_{\rm pl}$, $q=2$, we need $N\sim 20$ copies to match the observed level of perturbations.
We should also note that possible one-loop contributions from the matter sector to the inflaton potential are suppressed, at least by a factor of $q^{-N}/16\pi^2$, and therefore such contributions are fully under control. 

Another issue the clockwork can be useful for in large field models is the one of super-planckian field excursion. This was already noticed in Ref. \cite{kr}. If the scalar perturbations are ascribable to only one scalar degree of freedom,  then

\begin{equation}
\label{k}
\frac{\Delta\phi}{\mpl}\simeq \left(\frac{r}{2\times 10^{-2}}\right)^{1/2},
\end{equation}
where $r$ is the so-called tensor-to-scalar ratio. A future detection of   gravitational waves requires in general
variation of the inflaton field of the order of the Planck scale \cite{lythgrav}. This would pose a  problem as  
 slow-roll models of inflation disregard the possible presence of 
higher-order operators with powers of $(\phi/\mpl)$. However, super-Planckian field excursions can be mimicked while preserving the regime of  renormalizable four-dimensional field theory by the clockwork mechanism. Indeed,  the slow-roll of the  inflaton field over  super-planckian field values corresponds to  a clockwork of phase-rotations of the fields $N+1$  copies of fields $\pi_i$ with a U(1)$^{N+1}$  global symmetry whose effective decay constant is amplified with respect to the original one by a factor $q^N$.

\subsection{Hybrid models of inflation}
To illustrate the advantage of the clockwork mechanism in terms of efficiently producing small couplings during inflation, 
let us consider the hybrid model of inflation \cite{hybrid,stewart} with $N+1$ copies  of the fields $\pi_i$ and and extra field $\Phi$

\beq
\label{a}
V(\pi_1,\cdots,\pi_N,\Phi)=\frac{M^2}{2}\sum_{i=0}^{N-1}\left(\pi_i-q \,\pi_{i+1}\right)^2+V_0\left(1-\frac{\Phi}{M}\right)+\frac{1}{4}\lambda\pi_N^2\Phi^2+\cdots.
\eeq
The dots represent one or more additional terms, which give the potential  a minimum at which it vanishes but play no role during inflation. By performing the standard diagonalization of the clockwork mechanism for the  mass squared term in Eq. (\ref{a}) and working at energies
much smaller than the Hubble rate $H$, the potential (\ref{a}) for the lightest eigenstate $\phi_0$ is reduced to

\beq
V(\phi_0,\Phi)=V_0\left(1-\frac{\Phi}{M}\right)+\frac{\lambda_0}{4}\lambda\phi_0^2\Phi^2+\cdots,\,\,\,\,\lambda_0=\frac{\lambda}{q^{2N}}.
\eeq
For suitable choices of the parameters,  inflation takes place with the field $\Phi$ held at the instantaneous minimum, leading to a potential

\beq
V(\phi_0)=V_0\left(1-\frac{V_0}{\lambda^2M^2\phi_0^2}\right).
\eeq
Imposing the condition (\ref{cobe}) gives \cite{lr}

\beq
10^{-12}\simeq \lambda_0\frac{V_0^{1/2} M}{\mpl^3},
\eeq
and one could explain a small coupling $\lambda_0$ with the clockwork mechanism even though all the other mass scales in the problem are
of the order of the Planck scale.

\subsection{Small field inflation}
In small field models of inflation the problem is to have a flat enough potential close to the origin

\beq
V=V_0-\frac{1}{2}m^2\phi^2+\cdots. 
\eeq
As the spectral index of the scalar perturbations is given by \cite{lr} $1-n=2\mpl^2m^2/V_0\sim 0.04$, one needs $m^2\sim 10^{-1}H^2$ to be in agreement with the
observations. To produce a potential suitable for small field inflation, in the clockwork scenario it is enough to couple the pion $\pi_N$ to fermions charged under some strong group. Below the confinement scale the lightest mode acquires a potential of the form used in natural inflation \cite{natural}

\beq
V(\phi_0)=\Lambda^4\cos\frac{\phi_0}{q^N f},
\eeq
in such a way that, expanding around the maximum of the potential we obtain

\beq
V_0=\Lambda^4,\,\,\,\, m^2=\frac{\Lambda^4}{q^{2N} f^2}
\eeq
and the condition $m^2\ll H^2$ requires

\beq
f^2\gg \frac{\mpl^2}{q^{2N}}.
\eeq
This condition can be easily satisfied if  $f\ll \mpl$ even  for moderate values of $N$\footnote{In this respect, the clockwork mechanism applied to natural inflation is reminiscent of the aligned mechanism \cite{aligned1,aligned2,aligned3}. However, alignment does require a hierarchy of parameters which is naturally achieved through clockwork.}. Furthermore, the normalization of the density perturbations (\ref{cobe}) imposes

\beq
\frac{V_0^{1/2}}{\mpl^2}\simeq 5.4\cdot 10^{-4}\frac{1-n}{2}e^{{\cal N}_{\rm e}(1-n)/2}\frac{\phi_{\rm e}}{\mpl},
\eeq
where ${\cal N}_{\rm e}$ is the number of $e$-folds till the end of inflation and $\phi_{\rm e}$ is the value of the inflaton field $\phi_0$ when inflation ends.
Since the typical scale for $\phi_{\rm e}$ is $q^N f\gg \mpl$, one sees that the clockwork can allow sizeable $\Lambda$.

\subsection{Starobinsky inflation}
Another illustrative example of the efficiency of the clockwork mechanism  is provided  by the so-called Starobinsky model of inflation \cite{star}. Let us consider $N+1$ copies of  GR with action

\beq
S=\sum_{i=0}^{N}\frac{M_i^2}{2}\int \d ^4x \sqrt{-g_i}\,R(g_{i\,\mu\nu})+
\frac{1}{12\alpha^2} \int \d ^4x \sqrt{-g_N}R^2(g_{N\,\mu\nu}), \label{st}
\eeq
where we have added a quadratic curvature term for the $N$-th term whose strength is parametrized by a dimensionless parameter $\alpha$. It is known that in $(R+R^2)$-theory, there is a massive scalar mode on top of the  gravitons which can be uncovered with the usual methods,  leading   to the action 

\begin{eqnarray}
S&=&\sum_{i=0}^{N}\frac{M_i^2}{2}\int \d ^4x \sqrt{-g_i}\,R(g_{i\,\mu\nu})\nonumber \\
&+&
\int \d^4 x \sqrt{-g_N}\left[
-\frac{1}{2}(\partial \phi_N)^2-\frac{3}{4}\alpha^2 M_N^4\left( 1-e^{-\sqrt{\frac{2}{3}}\, \frac{\phi_N}{M_N}}\right)^2\right].  \label{st1}
\end{eqnarray}
For this theory, as  shown in Ref. \cite{gm}, there is a massless graviton with a corresponding Planck mass, which in the large $N$-limit is 
\begin{eqnarray}
M_{\rm pl}^2=q^N M_N^2.
\end{eqnarray}
Therefore, the action for the massless graviton and the scalar becomes 
 
 \begin{eqnarray}
S=\int \d ^4x \sqrt{-g}\left\{\frac{M_{\rm pl}}{2}R
-\frac{1}{2}(\partial \phi_N)^2-\frac{3}{4}\alpha^2 M_{\rm pl}^4q^{-4N}\left( 1-e^{-\sqrt{\frac{2}{3}}\, \frac{\phi_N q^N}{M_{\rm pl}}}\right)^2 \right\}. \label{st1}
\end{eqnarray}
During inflation $\phi_N$ takes large values and  the dynamics is dominated by the vacuum energy
\begin{eqnarray}
V=\frac{3}{4}\alpha^2 M_{\rm pl}^4\, q^{-4N}.
\end{eqnarray}
In  order to get the correct normalization (\ref{cobe})   one needs 
$q^{-2N}\approx 10^{-5} $ for  $\alpha={\cal O}(1)$ . For instance, for $q=2$, the moderate value
$N= 8$ is required. In addition, the tensor-to-scalar ratio is $r=(12 q^{-2N}/{\cal N}_{\rm e}^2)$. Hence, the tensor modes are 
suppressed  by an extra  factor of $q^{-2N}=10^{-5}$, leaving to no room for tensor modes in the clockwork Starobinsky inflation model.

\subsection{The clockwork and the generation of  perturbations from light fields other than the inflaton}
Even though the inflationary paradigm is by itself quite elegant and simple,  the mechanism
giving rise to the  adiabatic cosmological perturbations   is far from being 
established. It is fair to say  that  we do not know at present
what is the   source of the scalar perturbations during inflation, the inflaton field itself or some other field.
The total curvature perturbation $\zeta$ might not be  a constant (in time) on super-Hubble scales and 
  changing  on arbitrarily large scales due to a non-adiabatic
pressure perturbation    which may be 
due to  extra scalar degrees of freedom.

For instance, in  the curvaton 
mechanism
\cite{curvaton1,LW}  the  curvature perturbation
is generated from an initial isocurvature perturbation associated to the
quantum fluctuations of a light scalar field $\sigma$,  
the curvaton, whose energy density is not dominant  during inflation. The 
curvaton isocurvature perturbation becomes  the adiabatic
one once the curvaton decays into radiation  after the end 
of inflation. 

During inflation 
a flat spectrum is produced in the curvaton field

\beq
\delta\sigma = \left(\frac{H}{2\pi}\right)_*.
\eeq 
After 
 inflation, 
the curvaton field starts oscillating  during some radiation-dominated era,
causing its energy density to increase and 
converting the initial isocurvature into curvature 
perturbation $\zeta$. 

The curvaton mechanism works as long as the curvaton can be quantum mechanically excited during inflation. This requires
that the mass $m$ of the curvaton field is much smaller than the Hubble rate during inflation, $m\ll H$. This poses a problem as 
non-renormalizable couplings between the inflaton and the curvaton are expected to generate ${\cal O}(H^2)$ correction to the mass squared of the
curvaton. For instance, in supergravity one has
 corrections to the K\"{a}hler potential of the curvaton of the form

\be
\delta {\cal K}\supset\int\,{\rm d}\theta\, \frac{\phi^\dagger\phi}{M_{\rm pl}}\, \sigma^\dagger\sigma\supset {\cal O}(1) H^2\sigma^\dagger\sigma,
\ee
where $\phi$ is the inflaton field. The clockwork provides a possible solution to this problem.
Suppose that there are $N+1$ copies of curvatons, let us call them again $\pi_i$ with potential

\beq
\label{aa}
V(\pi_1,\cdots,\pi_N)=\frac{M_1^2}{2}\sum_{i=0}^{N-1}\left(\pi_i-q \,\pi_{i+1}\right)^2.
\eeq
This potential has the usual shift symmetry $\pi_i\to\pi_i+1/q^i$. Like in the construction of Ref. \cite{kr}, this shift symmetry is a manifestation of the fact that the $\pi_i$'s are indeed pseudo Nambu-Goldstone bosons of a U(1)$^{N+1}$ global symmetry. Since gravity is expected to break such a global symmetry, one expects
corrections to the mass of the form

\beq
\sum_{i,j=0}^{N} c_{ij} H^2\pi_i\pi_j=c_{NN}H^2 \pi_N^2+\cdots,
\eeq
where the $c_{ij}$ are ${\cal O}(1)$ coefficients and we have supposed that the vacuum energy is located at the $N$-th site.
Upon diagonalizing the mass matrix (\ref{aa}) one finds that the lightest eigenstate $\sigma_0$ receives corrections to its mass squared suppressed at least by  $1/q^N\ll 1$, which is enough to obtain a light curvaton during inflation.

\section{Clockwork inflation: the  five-dimensional continuum perspective}
In this section we investigate the  clockwork inflationary scenario from the continuum limit point of view. We imagine that the vacuum energy driving inflation is located at one of the two branes and, as a result, each fifth dimensional section is inflating with constant Hubble rate.
Let us start with  the five-dimensional  action \cite{little,gm}
\begin{eqnarray}
{\cal S}&=&\int \d ^5x\sqrt{-g}\left\{\frac{M_5^3}{2}
\left(R-\frac{1}{3}\partial_M S\partial^M S+4k^2 e^{-\frac{2}{3}S}\right)+{\cal L}_{\rm{bulk}}\right\}+{\cal S}_{\rm brane}, \label{s1}
\end{eqnarray}
where 
\begin{eqnarray}
{\cal S}_{\rm brane}=
-\sum_a \int \d^4 x \sqrt{h}\left(\frac{M_5^3}{2}[{\cal K}]+e^{-\frac{1}{3}S}{\cal L}_{\rm{brane}}\right). 
\end{eqnarray}
${\cal L}_{\rm{bulk}}$ is a bulk matter action, ${\cal L}_{\rm{brane}}$ is the brane action and $[{\cal K}]={\cal K}^+-{\cal K}^-$ is the jump of the trace of the extrinsic curvature 
\begin{eqnarray}
{\cal K}_{\mu\nu}=h^M_\mu h^N_\nu\nabla_M n_N,
\end{eqnarray}
defined in terms of the normal to the branes $\Sigma_a$ located at positions $x_a$ and the projection tensor $h^M_\mu$. The field equations that follows from the action (\ref{s1}) are
\begin{eqnarray}
&&R_{MN}-\frac{1}{2}g_{MN}R=\frac{1}{M_5^3} T_{MN},\label{ein1}\\
&&\nabla^2 S-4k^2 e^{-\frac{2}{3}S}=0, \label{es}
\end{eqnarray}
supplemented by  the Israel matching conditions for the branes $\Sigma_a$ 
\begin{eqnarray}
&&[{\cal K}_{\mu\nu}]_{\Sigma_a}=
\frac{1}{M_5^3}\left(T^{(a)}_{\mu\nu}-\frac{1}{3}g_{\mu\nu} T^{(a)}\right)\Big{|}_{\Sigma_a}, \nonumber \\
&&[n^M\partial_M S]_{\Sigma_a}=\frac{3}{M_5^3}\frac{\partial}{\partial S}\left(e^{-\frac{1}{3}S} {\cal L}_{\rm{brane}}\right)\Big{|}_{\Sigma_a} ,
\end{eqnarray}
and $T^{(a)}_{\mu\nu}$ is the four-dimensional brane energy-momentum tensor.
We are looking now for solutions of the form
\begin{eqnarray}
\d s^2=e^{2\sigma(y)}\left(\d y^2+\d s_{4}^2\right), ~~~S=S(y), \label{met}
\end{eqnarray}
where 
\begin{eqnarray}
\d s_{4}^2=g^{\rm dS}_{\mu\nu}\d x^\mu \d x^\nu=\frac{1}{H^2 \eta^2}\left(-\d \eta^2+\d \vec{x}^2\right),
\end{eqnarray}
is a four-dimensional de Sitter metric. 
The equations of motion are then 
\begin{eqnarray}
&&36(\sigma'^2-H^2)-S'^2=12k^2 e^{-\frac{2}{3}S+2\sigma},\label{e2}\\
&&\sigma''-\sigma'^2+H^2+\frac{1}{9}S'^2=0, \label{e3}\\
&&S''+3\sigma' S'=4k^2 e^{-\frac{2}{3}S+2\sigma}\label{e4}.
\end{eqnarray}
In addition, the Israel matching conditions along the branes $\Sigma_a$ located at $y=y_a$ with tension ${\cal L}_{\rm{brane}}^{(a)}=V_a$ for the metric (\ref{met}) give 
\begin{eqnarray}
&&\big{[}\sigma'\big{]}_{\Sigma_a}=
-\frac{1}{6M_5^3}e^{\sigma_a-\frac{1}{3}S_a} V_a, \nonumber \\
&&\big{[}S'\big{]}_{\Sigma_a}=-\frac{1}{2M_5^3}e^{\sigma_a-\frac{1}{3}S_a} V_a,  \label{isra}
\end{eqnarray}
where  $\sigma_a=\sigma(y_a),~S_a=S(y_a)$.
 The scalar field equation (\ref{e4}) is not independent as it is  connected to the Bianchi identity. So only Eqs. (\ref{e2}) and (\ref{e3}) are independent  from where, 
eliminating $S'^2$ we get the system of equations
\begin{eqnarray}
&&3\sigma''+9\sigma'^2-9 H^2=4 k^2 e^{-\frac{2}{3}S+2\sigma}\label{e22}, \\
&&\sigma''-\sigma'^2+H^2+\frac{1}{9}S'^2=0. \label{e33}
\end{eqnarray}
When $H=0$, the solution to Eqs. (\ref{e22}) and (\ref{e33}) is the linear dilaton solution
\begin{eqnarray}
\sigma=\sigma_0=\frac{2k}{3}y, ~~~S=S_0=2k y, \label{s0}
\end{eqnarray}
where the boundary condition $\sigma_0(0)=S_0(0)=0$ has been assumed.  However, the solution of Eqs. (\ref{e22}) and (\ref{e33}) for non-zero $H$ is not easy to be found. Nevertheless,   
by using Eq. (\ref{e22}) we can express the scalar $S$  
in terms of the wrap factor as
\begin{eqnarray}
 e^{-\frac{2}{3}S}=\frac{3}{4k^2} e^{-2\sigma}\left(\sigma''+3\sigma'^2-3 H^2\right) \label{sig}.
\end{eqnarray}
By taking the derivative of Eq. (\ref{e33})  and comparing with Eq. (\ref{e22}), we can completely decouple the scalar $S$ and   we find that 
$\sigma$ satisfies the equation the third order equation
\begin{eqnarray}
\Big{[}2\sigma'\left(3 H^2-3 \sigma'^2+2\sigma''\right)+\sigma'''\Big{]}^2
+4\Big{(}H^2-\sigma'^2+\sigma''\Big{)}\Big{(}3H^2-3 \sigma'^2-\sigma''\Big{)}^2=0 . \label{e44}
\end{eqnarray}
Although we were not able to find an exact solution to Eq.(\ref{e44}) we can try to find a solution perturbatively in $H^2$. For this, we may write
\begin{eqnarray}
 \sigma\approx \sigma_0+H^2 \sigma_1
 \end{eqnarray} 
 and treat $\sigma_1$ as a first order  perturbation. We then find that $\sigma_1$ satisfies
 \begin{eqnarray}
 \sigma_1'''+2k\sigma_1''-2k=0, 
  \end{eqnarray}
and therefore, 
\begin{eqnarray}
 \sigma_1=\frac{1}{2}y^2+C_1 e^{-2k y}+C_3 y+C_2, \label{s1}
 \end{eqnarray} 
 where $C_{1,2,3}$ are integration constants.
 In addition, we find  that $S$  is given by 
 \begin{eqnarray}
 S\approx 2ky+\frac{3}{2}H^2\left[y^2+5 C_1 e^{-2ky}+y\left(2C_3-\frac{3}{k}\right)+2C_2-\frac{3C_3}{k}+\frac{3}{2k^2}\right].  \label{S1}
 \end{eqnarray}
 It is straightforward to verify that 
 $\sigma=\sigma_0+H^2\sigma_1$ and $S$ in Eq. (\ref{S1}) indeed satisfies the equations (\ref{e22}) and (\ref{e33}) to leading order in $H^2$.  
 With a compact fifth dimension  and two branes at $y=0$ and $y=\pi R$ with  brane action 
\begin{eqnarray}
{\cal S}_{\rm brane}=\int \d^4x \int_0^{\pi R}\d y\sqrt{-g}\, \frac{e^{-\frac{1}{3}S}}{\sqrt{g_{55}}}
\left[ \delta(y)V_0+\delta(y-\pi)V_\pi\right],  \label{branes}
\end{eqnarray}
we find from Eq. (\ref{isra}) that the discontinuities of $\sigma,S$ should satisfy 
\begin{eqnarray}
3[\sigma']_{0,\pi R}= [S']_{0,\pi R}.  \label{isra1}
\end{eqnarray}
The solutions that satisfy (\ref{isra1}) have $C_1=0$ and  if we take   $C_2=0$ and  $C_3=1/2k$
so that  $\sigma(0)=S(0)=0$, we get 
\begin{eqnarray}
 \sigma&=&\frac{2k}{3}|y|+\frac{1}{2}H^2\left(y^2+\frac{1}{k} |y|\right),\label{st}\\
 S&=& 2k|y|+\frac{3}{2}H^2\left(y^2+\frac{1}{k}|y|-\frac{3}{k}y \right).  \label{St}
 \end{eqnarray} 
  Note that if $S_\pi=S(\pi R)$, we have from Eq. (\ref{St}) 
\begin{eqnarray}
S_\pi=2k\pi R+\frac{3}{2}H^2\left(k^2 \pi^2 R^2-\frac{2}{k}\pi R \right),
\end{eqnarray}
which fixes the radion field $R$ in terms of the boundary value of the dilaton $S$ at $y=\pi R$. The latter can be  determined by a potential for example of the form

\beq
U(S)=\frac{M_5^2}{2}\left(S-S_\pi\right)^2.
\eeq
Such a potential  fixes the boundary value of the dilaton and consequently the value of $R$.
 In addition, the tensions $V_0$ and $V_\pi$ in Eq. (\ref{branes}) turn out to be
 \begin{eqnarray}
 V_0(H)=-\left(4k+\frac{3H^2}{k}\right)M_5^3, ~~~V_\pi(H)=\left(4k+\frac{3 H^2}{k}\right)e^{-\sigma(\pi R)+S(\pi R)/3} M_5^3. 
 \end{eqnarray}
At leading order in $H^2/k^2$, the extra vacuum energy driving inflation in the $y=\pi R$ brane\footnote{We are adopting the so-called $\pi$-frame here, where  the SM resides in the $y=\pi R$ brane,  as  opposed to  the 0-frame adopted in \cite{gm} for the continuum clockwork. The two frames are equivalent, and there is a well-defined transformation that connects the two frames. } is given by

\beq
U_\pi=e^{4\sigma(\pi R)-S(\pi R)/3} \left(V_\pi(H)-V_\pi(0)\right)=
\frac{3H^2}{k} e^{2k\pi R} M_5^3. \label{f1}
\eeq
We may write the last equation   as 
\begin{eqnarray}
H^2=\frac{U_\pi}{3M_{\rm pl}^2}, \label{fre}
\end{eqnarray}
which is just the standard Friedmann  law on the brane expected
with a stabilized dilaton to lowest order in $H^2$. In (\ref{fre}) 
 we have defined the  four-dimensional Planck mass in flat space-time as 
 \begin{eqnarray}
 M_{\rm pl}^2=\frac{M_5^3}{k}e^{2k\pi R}, \label{mmp}
 \end{eqnarray}
 which  coincides with the one found by dimensional reduction of the five-dimensional action reported in \cite{gm}. 
If the vacuum energy driving inflation comes from the $y=\pi R$ brane, the  scalar perturbations have  the same behavior at leading order in the slow-roll parameters as in the four-dimensional case \cite{trans1} and  we expect that  no detectable signature remains from the non-trivial geometry in the bulk\footnote{One can also check that  the five-dimensional  geometry is not gravitationally unstable due to the tachyonic growth of the  the conformal fluctuations of the four-dimensional metric \cite{trans3}.}.  The reason is the following.  Differently from the  tensor modes, which are genuinely five-dimensional free fields quantized in the
bulk,  scalar metric perturbations are generated by  the brane scalar field which  is quantized on
the brane and is  four-dimensional  in all regimes. The coupling between the inflaton field and the metric concerns  the
 long-wavelength limit  for which  the metric evolves as in the four-dimensional theory. Moreover, such a  coupling
is localized on the brane, and  no signature remains from the warped geometry in the
bulk. Corrections to this result are of the order of $-\dot H R^2=\epsilon (HR)^2$ as metric perturbations probe  the extra dimension on times scales $-H^2/\dot H$.
The same logic  is not valid for  tensor modes, as we now proceed to  discuss.

%
%
%

\section{Tensor modes in clockwork inflation}
In this section we study the behaviour of the tensor modes during a de Sitter stage in the clockwork set-up. The goal is to show that tensor modes are suppressed in this scenario with respect to the standard case, the reason being that the tensor modes feel the bulk by full strength and that during the de Sitter stage the latter is more warped.

We  start from the five-dimensional continuum perspective.
The advantage of using tensor modes  to probe the clockwork mechanism  is that there is a  dependence   only on the geometry, not on the microscopic models of inflation and stabilization of the extra fifth dimension. 

\subsection{Tensor modes: the five-dimensional continuum perspective}
As shown in Refs. \cite{trans1,trans2}, a massless  tensor mode is produced in braneworls scenarios  from inflation with the amplitude $H/\mpl$. The point is that   the effective Planck mass during inflation for  a curved de Sitter  braneworld  differs from that of the flat brane at low energies, due to a dependence on the Hubble rate $H$. Thus,  the amplitude of tensor modes  from inflation in a clockwork scenario  is usually different from a standard period of inflation without clockwork.

Let us indeed consider the equation of  motion of the graviton field in the continuum clockwork scenario starting from the factorized metric (\ref{met}). By decomposing the five-dimensional tensor mode as

\beq
h_{MN}(y,x^\rho)=e^{-3\sigma(y)} h_m(y) Q^{(m)}_{\mu\nu}(x^\rho),
\eeq
where $m$ is the eigenvalue corresponding to the Kaluza-Klein modes and the transverse-traceless conditions are imposed on 
$ Q^{(m)}_{\mu\nu}(x^\rho)$, one can show that that the eigenvalue problem can be written as \cite{trans1,trans2}

\begin{eqnarray}
&&-{\cal D}_{-}{\cal D}_{+} h_m(y)=m^2 h_m(y),\nonumber\\
{\cal D}_{\pm}&=&\partial_y\mp\frac{3}{2}\sigma'(y)=\partial_y
\mp\left[\frac{2k}{3}\theta(y)+\frac{H^2}{k^2}\left(y +2k\theta(y) e^{-2ky}\right)\right].
\end{eqnarray}
From this equation one immediately reads off the presence of the zero mode graviton with wavefunction

\beq
h_0(y)=e^{3\sigma(y)}.
\eeq
Furthermore, one can show that there is a mass gap for the other Kaluza-Klein modes of at least $3H^2/2$ in the mass squared. The key point is that
during inflation the effective-four dimensional Planck mass as seen by the massless tensor mode is different from the one in flat space since the
function $\sigma(y)$ in de Sitter, as shown in the previous section, 

\beq
 \sigma(y)=\frac{2k}{3}|y|+\frac{1}{2}H^2\left(y^2+\frac{1}{k} |y|\right)+{\cal O}\left(H^4 R^4\right) \label{s4}
\eeq
\begin{figure}[h!]
    \begin{center}
      \includegraphics[scale=.5]{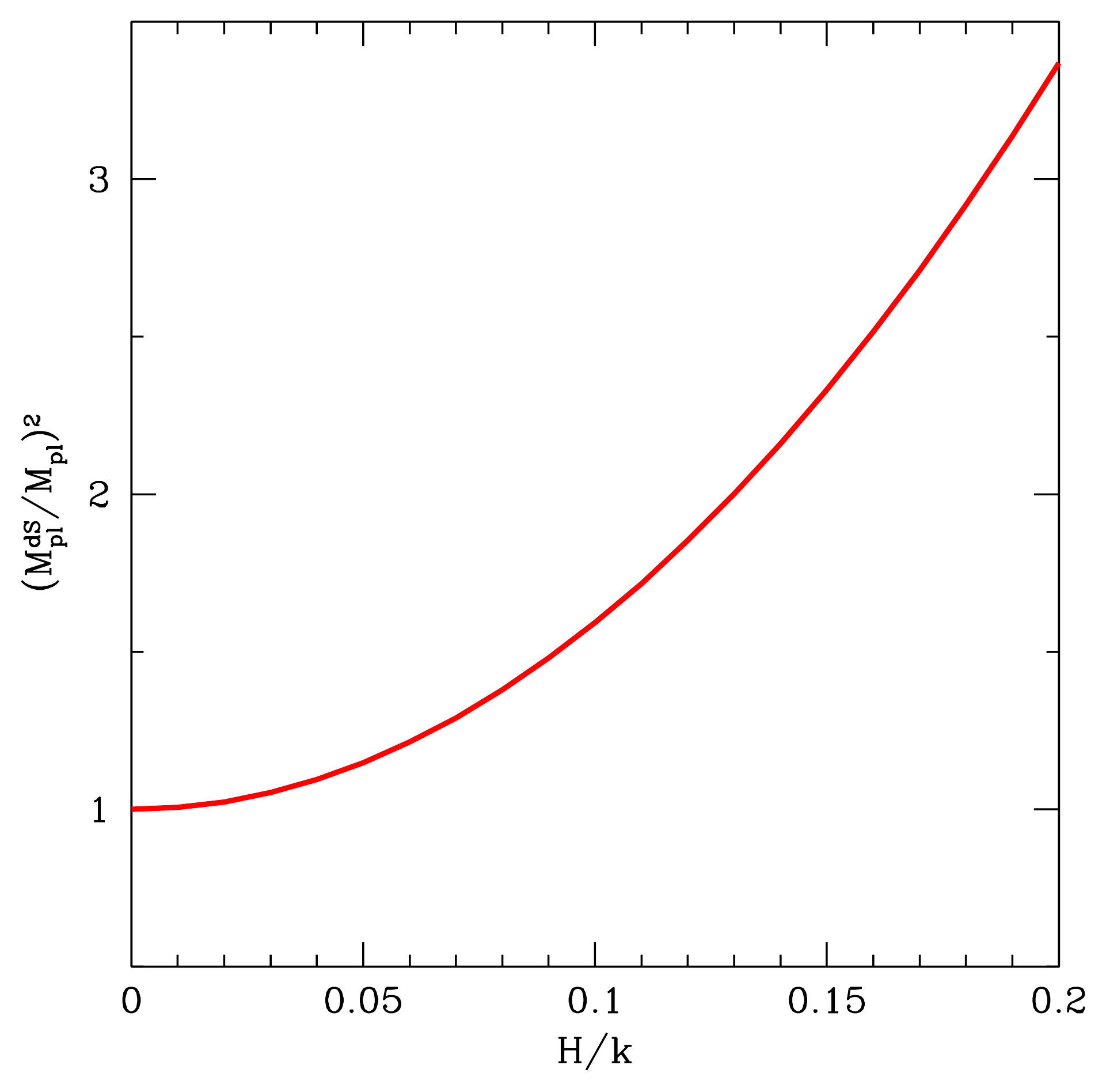}
    \end{center}
     \caption{\small The ratio of the Planckian mass during the de Sitter stage over the Planckian mass today as a function of $H/k$. Here we have taken $R= 2/k$.}  
\end{figure}
\noindent
is modified from the flat case. It is easy to see that the wrap factor $e^{\sigma(y)}$ is convex and  an increasing function of $H^2$.
 Therefore, the four-dimensional $\mpl$ defined as 
 \begin{eqnarray}
 \left(M^{\rm dS}_{\rm pl}\right)^2=2 M_5^3 \int_0^{\pi R} \d y\, e^{3\sigma}
 =\frac{M_5^3}{k}\left[e^{2k\pi R}\left(1+\frac{3}{2}H^2 \pi^2 R^2\right)-1\right],
 \end{eqnarray}
 is an increasing function of $H^2$, see Fig. 1. As the amount of tensor modes is proportional to $(H/\mpl^{\rm dS})^2$, this means that in the clockwork scenario the amount of tensor modes is reduced with respect to traditional case. Is this a negative point? Not necessarily so. There are many models of inflation which has been ruled out by the recent Planck data \cite{planck} as they produce a too large amount of tensor modes. Examples are the chaotic  large field model of inflation $\lambda\phi^4$  \cite{linde1} and  power-law inflation \cite{mat}. By embedding  these models into the  clockwork theory, these models become allowed again by current data.

\subsection{Tensor modes: the four-dimensional discrete perspective}
Let us analyze  now the clockwork graviton in de Sitter space from the four-dimensional perspective. 
In particular we assume  $N+1$ copies of metrics $g^{\mu\nu}_i$ describing $N+1$ copies of general relativity with their associated $N+1$  massless gravitons and Planck mass $M_i$. The gravitons $h^{\mu\nu}_i$ are fluctuations around the de Sitter metric $g^{\rm dS}_{\mu\nu}$ such that $g_{i\, \mu\nu}=g^{\rm dS}_{i\,\mu\nu}+h_{i\, \mu\nu}/M_i^2$. 
Clockworking will break the $N+1$ diffeomorphisms to a single 
diffeomorphic invariance corresponding to a single massless graviton. The  clockwork dynamics will  then be described  by the following action

\begin{eqnarray}
S&=&\int \d ^4x \sqrt{-\gamma}\left\{-\frac{1}{4} \sum_{i=0}^{N}
h^{\mu\nu}_i \Big{(}-\Box+2H^2\Big{)} 
 h_{i\mu\nu}-\frac{m^2}{4}
\sum_{i=0}^{N-1}\Big{(}h_i^{\mu\nu}-q_{i+1} h_{i+1}^{\mu\nu}\Big{)}^2\right\},  \label{act}
\end{eqnarray}
%
where $h_i^{\mu\nu}$ is transverse traceless $\nabla_\mu h_i^{\mu\nu}=0, ~
\gamma_{\mu\nu}h_i^{\mu\nu}=0$. For $q_{i+1}=0$, the action above describes $N+1$ massive gravitons on de Sitter in transverse traceless gauge \cite{Higuchi,dw}. Note that we have allow for  different $q$'s in the mass term ($q_1\neq q_2\ldots\neq q_N$) since the de Sitter metric is not flat. We will motivate  this choice by  deconstructing  the five-dimensional space where we will see that this choice is necessary when the background is not flat in general.
The clockwork theory described by the action (\ref{act}) is invariant now under 
the transformation
\begin{eqnarray}
h^{\mu\nu}_i\to h^{\mu\nu}_i+\frac{1}{Q_i} \left(\nabla^\mu\xi^\nu+\nabla^\nu\xi^\mu\right),~~~~Q_i=\prod_{j=1}^i q_i=q_1\cdots q_i, \label{dif}
\end{eqnarray}
where $\xi^\mu$ is a vector in de Sitter. Therefore, we expect a massless graviton in the spectrum, the existence of which can be verified by diagonalizing the mass matrix
\begin{eqnarray}
M^2_h = m^2
\begin{pmatrix}
1 & -q_1 & 0 & \cdots &  & 0 \cr
-q_1 & 
1+q_1^2 & -q_2 & \cdots 
&  & 0 \cr
0 & -q_2 & 1+q_2^2 & \cdots & & 0 \cr
\vdots & \vdots & \vdots 
& \ddots & &\vdots \cr
 & & & & 1+q_{N-1}^2 & -q_N \cr
 0 & 0 & 0 &\cdots & -q_N & q_N^2
\end{pmatrix} \, .
\label{dsmass}
\end{eqnarray}
It can easily be verified that the mass matrix $M^2$ has a zero eigenvalue corresponding exactly to the symmetry of Eq. (\ref{dif}). 
The action (\ref{act}) can be written then  as 
\begin{eqnarray}
S_=\int \d ^4x \sqrt{-\gamma}\left\{-\frac{1}{4}\sum_{i=0}^{N}\wh^{\mu\nu}_i \Big{(}-\Box+2H^2\Big{)} 
 \wh_{i\mu\nu}-\frac{1}{4}\sum_{i=1}^{N} 
 m_i^2\wh_{i\mu\nu}^2\right\}, \label{actd}
\end{eqnarray}
where $m_i^2$ are the non-zero eigenvalues of $M^2$ and 
\begin{eqnarray}
 \wh_i^{\mu\nu}=O^Th_i^{\mu\nu}, ~~~O^TM^2 O=\mbox{\rm diag}\Big{(}m_0^2=0,m_1^2, \ldots,m_{N-1}^2\big{)}.
 \end{eqnarray} 
We see that the theory described a massless graviton $\wh_0^{\mu\nu}$ and $N-1$ massive spin-2 states $\wh_i^{\mu\nu}, ~(i=1,\ldots,N-1)$. 

Let us now deconstruct the clockwork direction. The graviton fluctuations around a four-dimensional de Sitter background will be described by the
action
\begin{eqnarray}
 {\cal S}=-\frac{1}{4}\int \d^4 x\int_{-\pi R}^{\pi R} 
\d y\, e^{3\sigma}\bigg[\nabla_\lambda h^{\mu\nu} \nabla^\lambda h_{\mu\nu}+
 \partial_y h_{\mu\nu} \partial_y h_{\mu\nu}+
 2H^2 h_{\mu\nu}h^{\mu\nu}\bigg],
 \end{eqnarray} 
which after deconstruction turns out to be (after redefining $h_{\mu\nu}\to e^{-\frac{3}{2}\sigma}h_{\mu\nu}$)
\begin{eqnarray}
 {\cal S}&=&-\frac{1}{4}\int \d^4 x
\bigg[\sum_{i=0}^{N}\Big{(}\nabla_\lambda h_i^{\mu\nu} \nabla^\lambda h_{i\mu\nu}+
 2H^2 h_{i,\mu\nu}h_i^{\mu\nu}\Big{)}\nonumber \\
 &+&
\frac{1}{a^2} \sum_{i=0}^{N-1} e^{3\sigma(i)}\Big{(} 
e^{-\frac{3}{2}\sigma(i+1)} h_{i+1}^{\mu\nu}-e^{-\frac{3}{2}\sigma(i)} h_{i}^{\mu\nu}\Big{)}^2\bigg].
 \end{eqnarray} 
 Simple manipulations lead to the following  expression 
 
 \begin{eqnarray}
 {\cal S}&=&-\frac{1}{4}\int \d^4 x
\bigg[\sum_{i=0}^{N}\Big{(}\nabla_\lambda h_i^{\mu\nu} \nabla^\lambda h_{i\mu\nu}\!\!+\!\!
 2H^2 h_{i\mu\nu}h_i^{\mu\nu}\Big{)}\!\!+\!\!
\frac{1}{a^2} \sum_{i=0}^{N-1} \Big{(} 
 h_{i}^{\mu\nu}\!\!\!-e^{-\frac{3}{2}(\sigma(i+1)-\sigma(i))} h_{i+1}^{\mu\nu}\Big{)}^2\bigg]. 
\nonumber\\
 && \label{dec}
 \end{eqnarray} 
 For the warp factor $\sigma(y)$ given in Eq. (\ref{s4}), the deconstruction is carried out in the so-called 0-frame of the discrete clockwork, where the SM resides in the first site \cite{gm}. To go to the $N$-frame used here as well as in Ref. \cite{gm} where the Standard Model (SM) is localized at the $N$-th site in the discrete clockwork, we should change $\sigma\to -\sigma$ so that $q_i\to 1/q_i$. In this case we find that     

\beq
q_{i+1}=e^{\frac{3}{2}(\sigma(i+1)-\sigma(i))}\simeq e^{ka}e^{\frac{3}{4}a\frac{H^2}{k}+\frac{3}{4}a^2H^2(1+2i)}.
\eeq
Hence, 
the charge increases in going from the first site to the $N$-th site, and  (\ref{dec}) can be written as 
\begin{eqnarray}
 {\cal S}=-\frac{1}{4}\int \d^4 x
\bigg[\sum_{i=0}^{N}\Big{(}\nabla_\lambda h_i^{\mu\nu} \nabla^\lambda h_{i\mu\nu}\!\!+\!\!
 2H^2 h_{i\mu\nu}h_i^{\mu\nu}\Big{)}\!\!+\!\!
\frac{1}{a^2} \sum_{i=0}^{N-1} \Big{(} 
 h_{i}^{\mu\nu}\!\!\!-q_{i+1} h_{i+1}^{\mu\nu}\Big{)}^2\bigg]. \label{dec1}
 \end{eqnarray} 
 This  is  identical to Eq. (\ref{act}) which is motivated here as the deconstructed action along the clockwork direction. 
When the $y={\rm const.}$ sections are flat, $\sigma$ is a linear function of $y$ as in  Eq. (\ref{s0}) and hence  we get that 
$q_1=q_2\ldots=e^{ka}.$
However, in  the case in which the  $y={\rm const.}$  sections are not flat, and in particular de Sitter spaces like in the present setup, then we have that 
$q_1\neq q_2\ldots \neq q_N$ and the charge in the $N$-th site is larger  in de Sitter than it would be in Minkowski spacetime, explaining why the Planck scale is larger during inflation.

\section{Conclusions}
The clockwork   is an ingenious mechanism to generate large mass/coupling hierarchies in theories where no small parameters are present to start with. In this paper we have offered an handful number of examples of how the clockwork set-up may help to construct 
inflationary models with no fine-tuning. Interestingly, clockwork inflation predicts an  amount of tensor modes which is smaller than in standard scenarios with no clockwork. While this result is bad news for current and future efforts in detecting tensor modes in the B-mode polarization of the CMB, it is certainly good news for inflation model builders as many models of inflation prematurely ruled out by Planck
observations for their excessive tensor mode power spectrum, 
are now back to business.

\section*{Acknowledgments}
We thank G. Giudice and M. McCullough  for reading the draft of the paper, for discussions and  for providing  useful  suggestions. A.R. is supported by the Swiss National Science Foundation (SNSF), project {\sl Investigating the
Nature of Dark Matter}, project number: 200020-159223.


\end{document}